\newcommand{\CLASSINPUTtoptextmargin}{19mm}%
\newcommand{\CLASSINPUTbottomtextmargin}{43mm}%
\newcommand{\CLASSINPUTinnersidemargin}{12.9mm}%
\newcommand{\CLASSINPUToutersidemargin}{12.9mm}%
\documentclass[conference,10pt,a4paper]{IEEEtran}
\usepackage{amsmath}
\usepackage{times}
\usepackage{graphicx}
\usepackage{multirow}
\usepackage[none]{hyphenat}
\usepackage{float}
\usepackage{subfig}
\usepackage{iftex}

\usepackage{amsmath,amssymb,amsfonts}
\usepackage{algorithm, algorithmic}
\usepackage{subcaption}
\usepackage{graphicx}
\usepackage{balance}
\usepackage{textcomp}
\usepackage{xcolor}

\usepackage[backend=bibtex,defernumbers=true,sorting=none,style=ieee]{biblatex}%
\ifPDFTeX%
\usepackage{t1enc}
\usepackage{times}
\fi%
\ifXeTeX%
\usepackage{fontspec}
\fi%
\ifLuaTeX%
\usepackage{fontspec}
\fi%
\makeatletter

\def\@maketitle{\newpage
\bgroup\par\addvspace{0.5\baselineskip}\centering%
\ifCLASSOPTIONtechnote
   {\bfseries\large\@IEEEcompsoconly{\sffamily}\@title\par}\vskip 1.3em{\lineskip .5em\@IEEEcompsoconly{\sffamily}\@author
   \@IEEEspecialpapernotice\par{\@IEEEcompsoconly{\vskip 1.5em\relax
   \@IEEEtitleabstractindextextbox{\@IEEEtitleabstractindextext}\par
   \hfill\@IEEEcompsocdiamondline\hfill\hbox{}\par}}}\relax
\else
   \vskip0.2em{\EuMWtitlesize\ifCLASSOPTIONtransmag\bfseries\LARGE\fi\@IEEEcompsoconly{\sffamily}\@IEEEcompsocconfonly{\normalfont\normalsize\vskip 2\@IEEEnormalsizeunitybaselineskip
   \bfseries\Large}\@title\par}\vskip1.0em\par
   \ifCLASSOPTIONconference%
      {\@IEEEspecialpapernotice\mbox{}\vskip\@IEEEauthorblockconfadjspace%
       \mbox{}\hfill\begin{@IEEEauthorhalign}\@author\end{@IEEEauthorhalign}\hfill\mbox{}\par}\relax
   \else
      \ifCLASSOPTIONpeerreviewca
         {\@IEEEcompsoconly{\sffamily}\@IEEEspecialpapernotice\mbox{}\vskip\@IEEEauthorblockconfadjspace%
          \mbox{}\hfill\begin{@IEEEauthorhalign}\@author\end{@IEEEauthorhalign}\hfill\mbox{}\par
          {\@IEEEcompsoconly{\vskip 1.5em\relax
           \@IEEEtitleabstractindextextbox{\@IEEEtitleabstractindextext}\par\hfill
           \@IEEEcompsocdiamondline\hfill\hbox{}\par}}}\relax
      \else
         \ifCLASSOPTIONtransmag
           {\@IEEEspecialpapernotice\mbox{}\vskip\@IEEEauthorblockconfadjspace%
            \mbox{}\hfill\begin{@IEEEauthorhalign}\@author\end{@IEEEauthorhalign}\hfill\mbox{}\par
           {\vspace{0.5\baselineskip}\relax\@IEEEtitleabstractindextextbox{\@IEEEtitleabstractindextext}\vspace{-1\baselineskip}\par}}\relax
         \else
           {\lineskip.5em\@IEEEcompsoconly{\sffamily}\sublargesize\@author\@IEEEspecialpapernotice\par
           {\@IEEEcompsoconly{\vskip 1.5em\relax
            \@IEEEtitleabstractindextextbox{\@IEEEtitleabstractindextext}\par\hfill
            \@IEEEcompsocdiamondline\hfill\hbox{}\par}}}\relax
         \fi
      \fi
   \fi
\fi\par\addvspace{0.0\baselineskip}\egroup}

\def\EuMWtitlesize{\@setfontsize{\EuMWtitlesize}{24}{24pt}}
\def\EuMWauthorsize{\@setfontsize{\EuMWauthorsize}{11}{11pt}}
\def\EuMWaffilsize{\@setfontsize{\EuMWaffilsize}{10}{10pt}}
\def\EuMWcaptionsize{\@setfontsize{\EuMWcaptionsize}{9}{10pt}}
\def\EuMWbibsize{\@setfontsize{\EuMWbibsize}{8}{10pt}}

\def\@IEEEauthorblockNstyle{\EuMWauthorsize\@IEEEcompsocnotconfonly{\sffamily}\@IEEEcompsocconfonly{\large}}
\def\@IEEEauthorblockAstyle{\EuMWaffilsize\@IEEEcompsocnotconfonly{\sffamily}\@IEEEcompsocconfonly{\itshape}\@IEEEcompsocconfonly{\large}}
\def\@IEEEauthordefaulttextstyle{\EuMWauthorsize\@IEEEcompsocnotconfonly{\sffamily}\sublargesize}

\def\thebibliography#1{\section*{\refname}%
    \addcontentsline{toc}{section}{\refname}%
    \EuMWbibsize\@IEEEcompsocconfonly{\small}\vskip 0.3\baselineskip plus 0.1\baselineskip minus 0.1\baselineskip
    \list{\@biblabel{\@arabic\c@enumiv}}%
    {\settowidth\labelwidth{\@biblabel{#1}}%
    \leftmargin\labelwidth
    \advance\leftmargin\labelsep\relax
    \itemsep \IEEEbibitemsep\relax
    \usecounter{enumiv}%
    \let\p@enumiv\@empty
    \renewcommand\theenumiv{\@arabic\c@enumiv}}%
    \let\@IEEElatexbibitem\bibitem%
    \def\bibitem{\@IEEEbibitemprefix\@IEEElatexbibitem}%
\def\newblock{\hskip .11em plus .33em minus .07em}%
\ifCLASSOPTIONtechnote\sloppy\clubpenalty4000\widowpenalty4000\interlinepenalty100%
\else\sloppy\clubpenalty4000\widowpenalty4000\interlinepenalty500\fi%
    \sfcode`\.=1000\relax}

\long\def\@makecaption#1#2{%
\ifx\@captype\@IEEEtablestring%
\par\@IEEEtabletopskipstrut
\else
\@IEEEfigurecaptionsepspace
\fi
\setbox\@tempboxa\hbox{\normalfont\footnotesize {#1.}\nobreakspace\nobreakspace #2}%
\ifdim \wd\@tempboxa >\hsize%
\setbox\@tempboxa\hbox{\normalfont\footnotesize {#1.}\nobreakspace\nobreakspace}%
\parbox[t]{\hsize}{\normalfont\footnotesize\noindent\unhbox\@tempboxa#2}%
\else
\ifCLASSOPTIONconference \hbox to\hsize{\normalfont\footnotesize\hfil\box\@tempboxa\hfil}%
\else \hbox to\hsize{\normalfont\footnotesize\box\@tempboxa\hfil}%
\fi\fi
\ifx\@captype\@IEEEtablestring%
\@IEEEtablecaptionsepspace
\else
\fi}

\newlength\tablecaptiontotableskip
\newlength\figuretocaptionskip
\def\@IEEEfigurecaptionsepspace{\vskip\figuretocaptionskip\relax}%
\def\@IEEEtablecaptionsepspace{\vskip\tablecaptiontotableskip\relax}%

\def\abstract{\normalfont%
\@IEEEabskeysecsize\bfseries\textit{\abstractname}\,\bfseries\textit{---}\,%
\@IEEEgobbleleadPARNLSP}%

\def\IEEEkeywords{\normalfont%
\@IEEEabskeysecsize\bfseries\textit{\IEEEkeywordsname}\,\bfseries\textit{---}\,%
\@IEEEgobbleleadPARNLSP}%
\def\endIEEEkeywords{\relax\vspace{0.67ex}%
\par\if@twocolumn\else\endquotation\fi%
\normalsize\normalfont}%

\DeclareRobustCommand*{\EuMWauthorrefmark}[1]{\raisebox{0pt}[0pt][0pt]{\textsuperscript{#1}}}%
\def\@IEEEauthorblockNtopspace{0ex}
\def\@IEEEauthorblockAtopspace{1mm}
\def\tablename{Table}
\def\thetable{\arabic{table}}
\def\IEEEkeywordsname{Keywords}
\def\subsubsection{\@startsection{subsubsection}{3}{\z@}{1.5ex plus 1.5ex minus 0.5ex}%
{0.7ex plus .5ex minus 0ex}{\normalfont\normalsize\itshape}}%
\newlength{\CPheadmatchindent}%
\def\@seccntformat#1{\hbox to\CPheadmatchindent{\csname the#1dis\endcsname}\hskip 0.1em \relax}
\IEEEilabelindentA \parindent
\IEEEilabelindent \IEEEilabelindentA
\IEEEelabelindent \parindent
\IEEEdlabelindent \parindent
\IEEElabelindent \parindent
\makeatother

\begin{document}
\raggedbottom
%
%
%
\title{N78 Frequency Band Modular RIS Design and Implementation}
%
%
\author{
\IEEEauthorblockN{%
Sefa Kayraklık\EuMWauthorrefmark{$\circ$}, 
Recep Baş\EuMWauthorrefmark{$\bullet$1},
Hasan Oğuzhan Çalışkan\EuMWauthorrefmark{$\ast$}, 
Samed Şahinoğlu\EuMWauthorrefmark{$\ast$}, 
Sercan Erdoğan\EuMWauthorrefmark{$\bullet$},\\
İlhami Ünal\EuMWauthorrefmark{$\bullet\diamond $2},
İbrahim Hökelek\EuMWauthorrefmark{$\circ$},
Kıvanç Nurdan\EuMWauthorrefmark{$\ast$},
Ali Görçin\EuMWauthorrefmark{$\circ$}
}
\IEEEauthorblockA{%
\EuMWauthorrefmark{$\circ$}Communications and Signal Processing Research (HISAR) Lab., TUBITAK BILGEM, Kocaeli, Türkiye\\ 
\EuMWauthorrefmark{$\bullet$}Millimeter Wave and Terahertz Technologies Research Lab. (MİLTAL), TUBITAK MAM, Kocaeli, Türkiye\\
\EuMWauthorrefmark{$\ast$}Embedded Systems and Digital Design Department, TUBITAK BILGEM, Kocaeli, Türkiye\\
\EuMWauthorrefmark{$\diamond $}now with School of Electrical and Electronic Engineering, University College Dublin, Dublin, Ireland\\
\{name.surname\}@tubitak.gov.tr \EuMWauthorrefmark{1}recepbas.iu@gmail.com \EuMWauthorrefmark{2}ilhami.unal@ucd.ie\\
}
}
%
\maketitle
%
%
%
{\renewcommand\thefootnote{}\footnotetext{Proceedings of the 55th European Microwave Conference (© 2025 EuMA)}}
\begin{abstract}
Reconfigurable intelligent surface (RIS), capable of dynamically controlling wireless propagation characteristics using reflecting antenna elements, is a promising technology for enhancing signal coverage and improving end-user connectivity in next-generation wireless networks. This paper presents a complete design flow of a modular RIS prototype operating at the n78 frequency band, starting from the simulations to the prototype development and testing. An RIS prototype includes one master and up to sixteen slave blocks, each of which has an identical hardware structure with $8\times 8$ reflecting surface elements and a controller board. The phase shift response of each unit element is controlled with a PIN diode to form a $180^\circ$ phase difference between the ON and OFF states. The measurement experiment using two RIS blocks, horn antennas, and a vector network analyzer showed that the improvement of the received signal power is more than $15$ dB across the n78 frequency band for a given placement. 
\end{abstract}

\section{Introduction}
Reconfigurable intelligent surfaces (RISs) are a promising technology for next-generation wireless communication systems due to their inherent capability of dynamically controlling the wireless propagation characteristics \cite{basar2019wireless}. RISs consisting of configurable, energy-efficient, and low-cost unit surface elements are anticipated to be an important element of future wireless communication networks to enhance signal coverage and improve end-user connectivity \cite{wen2024shaping}. Furthermore, the RIS's multifunction capabilities, encompassing communication, sensing, localization, control, and computing, make the RIS one of the key enablers in next-generation services and applications, such as integrated sensing and communication, non-terrestrial networks, vehicle-to-everything, and so on \cite{katwe2024overview}. Designing and manufacturing RIS prototypes are of paramount importance to validate their effectiveness in these advanced use cases and provide valuable contributions to the standardization bodies.

In \cite{araghi2022reconfigurable}, an RIS prototype, which consists of $2430$ reflecting elements operating at $3.5$ GHz, is designed and developed to demonstrate the beam steering capability to the blind spots by continuous phase shift adjustment of the incident signals. The authors in \cite{fara2022prototype} propose a phase and amplitude response model of the reflecting elements, characterize an RIS prototype with continuous phase shift tuning, and implement a codebook design for accurate beam focusing. In \cite{trichopoulos2022design}, a low-power, single-layer, PIN diode-based RIS prototype operating at $5.8$ GHz is presented, and the RIS's beamforming performance in real-world outdoor environments is demonstrated by improving the signal coverage in obstructed scenarios. Another RIS prototype \cite{rains2023high} with a 3-bit column-drive control mechanism is developed, and its performance in enhancing received signal power is evaluated in various indoor deployments. Dual-polarized 2-bit phase shift resolution RIS prototypes are developed with a high aperture efficiency \cite{yin2024design} and a high cross-polarization discrimination \cite{zhu2024dual}. 

In this paper, we design and manufacture a modular RIS prototype operating at the n78 frequency band, in which multiple blocks can be combined to obtain large-sized reflecting surfaces. First, the initial design of the unit reflecting element, consisting of four metallic layers and three dielectric substrate layers, is done by utilizing the CST Microwave Studio. The phase shift response of each unit element is controlled with a PIN diode to form a $180^\circ$ phase difference between the PIN diode ON and OFF states. Next, four RIS blocks ($2\times 2$), each consisting of $64$ elements and forming a total of $256$ elements in a $16 \times 16$ array with $41$ mm spacing in both rows and columns, are designed on the CST Microwave Studio, where the beamforming range of up to  $45^\circ$  at the $3.7$-$3.8$ GHz frequency band is achieved. To validate the design of the unit reflecting element before the RIS prototype development, one unit cell is fabricated, and its reflection parameter (\(S_1\)\(_1\)) is measured through the waveguide measurement. 
Next, an RIS prototype with one master and up to sixteen slave blocks, each of which has an identical hardware structure with $8\times 8$ unit elements, is manufactured. The master RIS block controls power distribution and sends configuration data to all connected slave blocks. A microcontroller (MCU) supporting Wi-Fi wireless communication, as well as I2C, UART, and USB interfaces, is used to facilitate self-configuration and inter-block communication. The magnitude of measured \(S_1\)\(_1\) parameters is over $-3$ dB when the PIN diode state is ON and OFF, and the difference of unwrapped phase values between the ON and OFF states is almost in the range of $180 \pm 20$ degrees for $3.7$-$3.8$ GHz. Finally, the measurement experiment for two RIS blocks demonstrates more than $15$ dB improvement over n78 frequency band. 

\section{Reconfigurable Intelligent Surface Design}

\subsection{Unit Reflecting Element Design}

\begin{figure}[t]
\centering
    \includegraphics[width=0.41\textwidth]{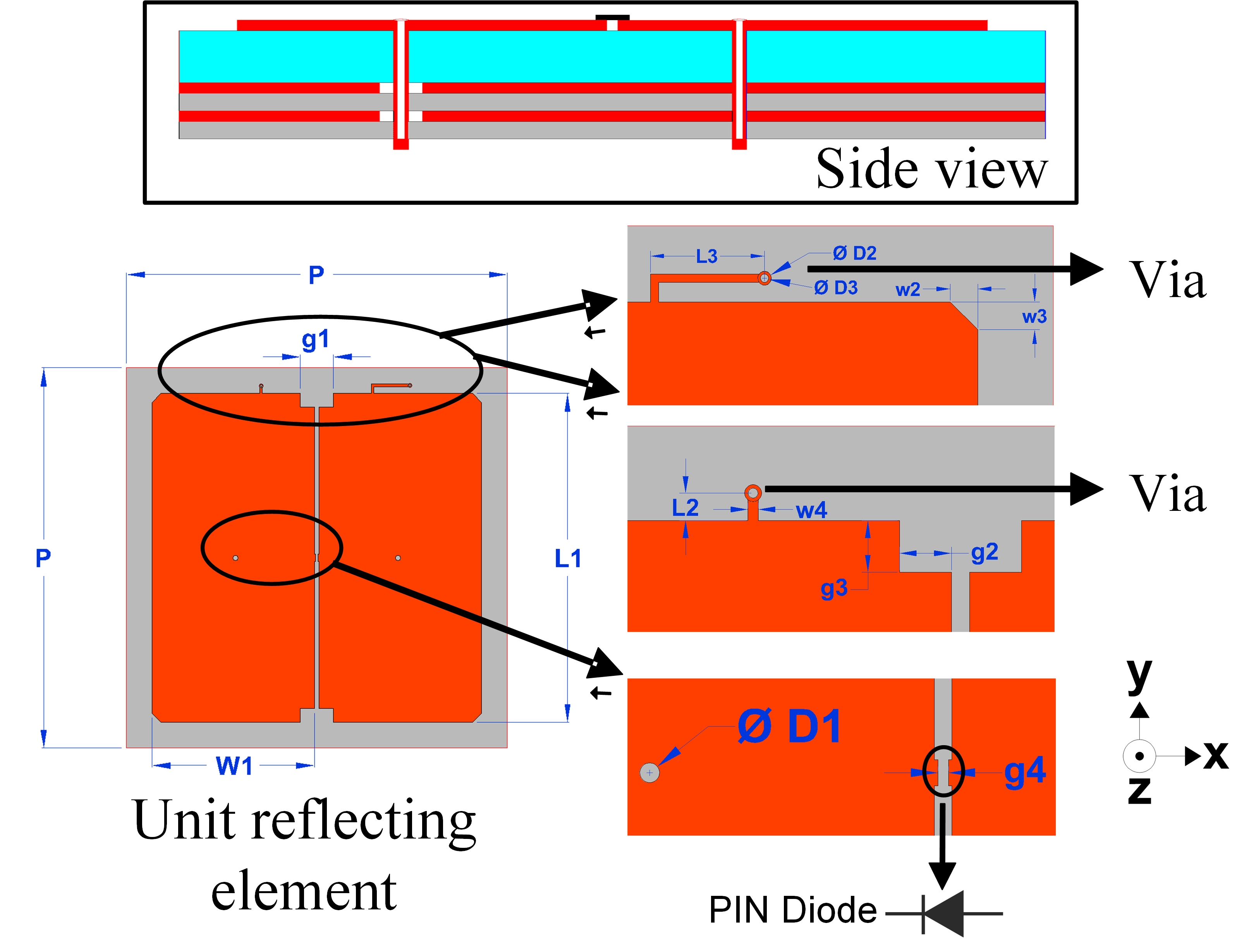}
\caption{Side view and top view of the reflecting element (P = 41 mm,     L1 = 35.5mm, W1 = 17.49mm, g1 = 3.525mm, L2 = 0.8mm, L3 = 4.15mm, w2 = w3 = 1mm, w4 = 0.3mm, g2 = g3 = 1.5mm, g4 = 0.31mm, D1 = 0.6mm, D2 = 0.5 mm, D3 = 0.3 mm) }
\label{fig:ris-sidetoplayer}\vspace{-10pt}
\end{figure}

\begin{figure}[t]
\centering
    \includegraphics[width=0.31\textwidth]{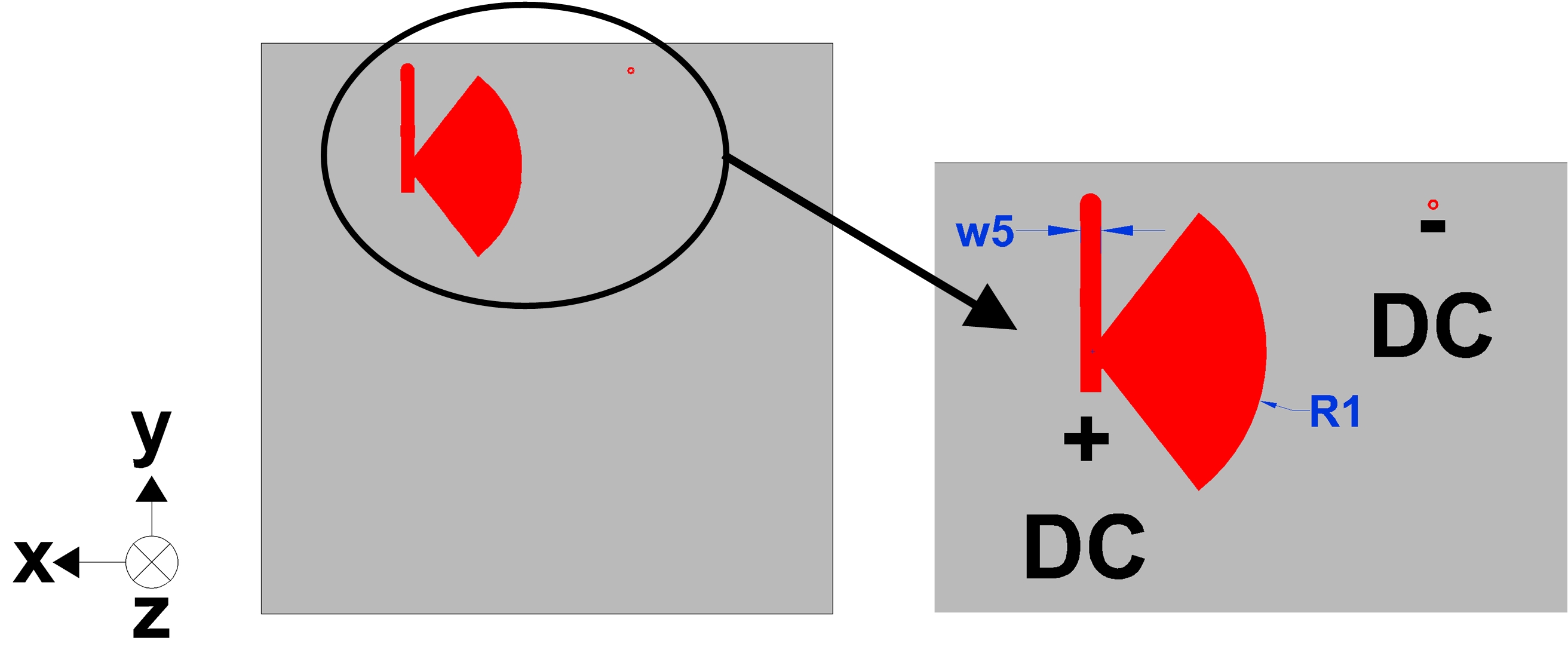}
\caption{Bottom view of the reflecting element (w5 =0.95 mm, R1 = 8.13mm)}
\label{fig:ris-bottomlayer}\vspace{-10pt}
\end{figure}

The structure consists of four metallic layers and three dielectric substrate layers. On the top metallic layer, two patches with a circular aperture are connected with a PIN diode. Also, two microstrip line extensions for vias are added to the patches for DC biasing of the PIN diode. The second and third metallic layers are used as RF ground and DC ground, which are fully covered with copper except for one of the bias via’s apertures, respectively. At the bottom layer, an RF choke is added to the bias line in order to prevent electromagnetic wave leakage to the bias circuit. The side and top views of the unit reflecting element are shown in Fig. \ref{fig:ris-sidetoplayer}, and the bottom view of the unit reflecting element is illustrated in Fig. \ref{fig:ris-bottomlayer}. The copper thickness of all metallic layers is $0.035$ mm. The first dielectric layer is F4BM with a $2.65$ dielectric constant, a $0.005$ loss tangent, and a thickness of $3$ mm. The remaining dielectric layers are FR4 with $4.4$ dielectric constant, $0.025$ loss tangent, and $0.5$ mm thickness.

\begin{figure}[t]
\centering
    \includegraphics[width=0.5\textwidth]{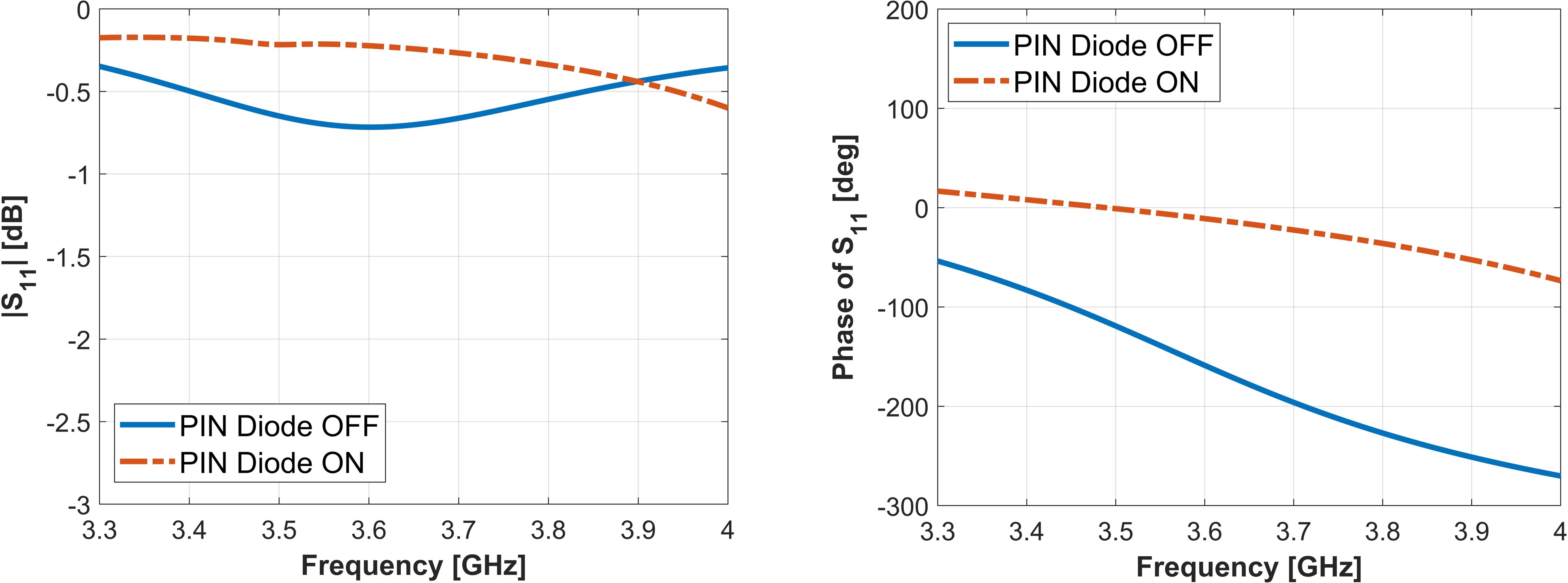}
\caption{Simulation results of reflection parameter \(S_1\)\(_1\) for one unit element}
\label{fig:sim-unitcellResult}\vspace{-10pt}
\end{figure}

The initial design of the unit reflecting element is prepared on the CST Microwave Studio, and the design is completed using parameter optimization. The phase shift response of each unit element is controlled with a PIN diode to form a $180^\circ$ phase difference between the PIN diode states in one polarization. The targeted operating frequency is $3.75$ GHz. 

A low insertion loss PIN diode SMP1340-040LF from Skyworks is used for simulations. The equivalent circuit of the ON/OFF case of the PIN Diode can be found on the product datasheet. Reflection (\(S_1\)\(_1\)) parameters of the unit reflecting element for PIN diode’s ON and OFF case at normal incidence angle simulation are shown in Fig. \ref{fig:sim-unitcellResult}. The magnitude of \(S_1\)\(_1\) parameters are over $-3$ dB for both cases, and the unwrapped phase difference between the ON and OFF cases for the PIN diode is almost in the range of $180 \pm 20$ degrees for $3.7$-$3.8$ GHz frequency band up to $45$-degree incidence angle. 

\subsection{RIS Block Design}
\begin{figure}[t]
\centering
    \includegraphics[width=0.46\textwidth]{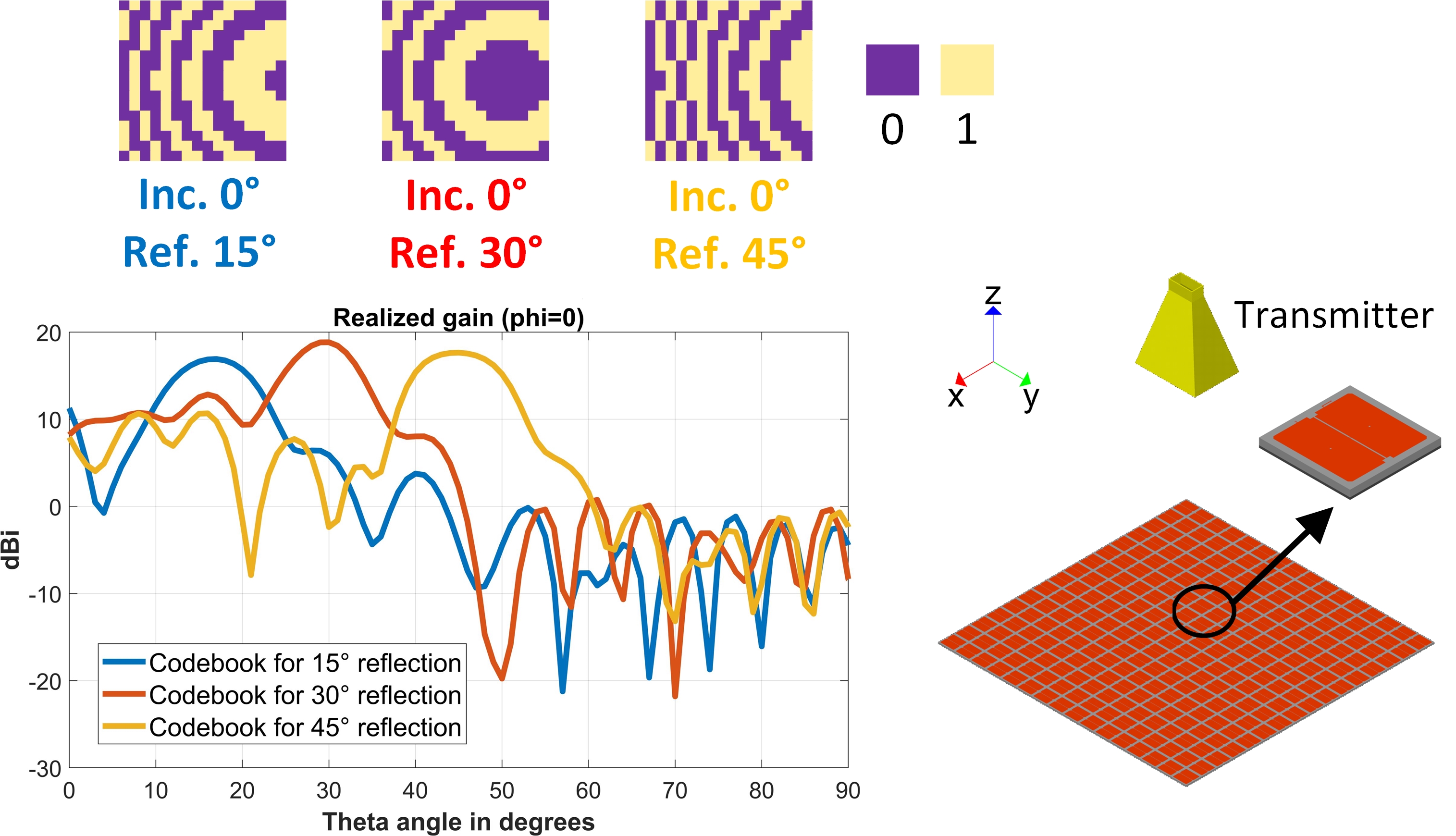}
\caption{Beamforming simulation results}
\label{fig:sim-blockResult}\vspace{-10pt}
\end{figure}

To simulate the beamforming performance, four RIS blocks ($2\times 2$), each consisting of $64$ elements and forming a total of $256$ elements in a $16\times 16$ array with $41$ mm spacing in both rows and columns, are prepared on the CST Microwave Studio. Fig. \ref{fig:sim-blockResult} illustrates the 3D view of the simulation setup, phase matrices (codebooks), and radiation pattern results for the center frequency of $3.75$ GHz. The distance between the antenna and RIS blocks is set at $800$ mm, and the incidence angle is kept at $0^\circ$. Three different phase matrices are prepared using array theory, and reflected angles are chosen as $15^\circ$, $30^\circ$, and $45^\circ$, respectively. The results show that the beamforming range of the RIS block is up to $45^\circ$ at the $3.7$-$3.8$ GHz. 

\subsection{Control Mechanism for RIS Block}

The RIS prototype system has been designed to be modular using identical and interchangeable RIS blocks, where multiple RIS blocks can be used in different layouts to obtain larger sized RISs. The proposed RIS system design consists of at least one master controller and up to sixteen slave controllers. Each RIS block includes a reflecting surface board and a controller module, which can be configured as master or slave. 

\begin{figure} [t]
    \centering
    \includegraphics[width=1\linewidth]{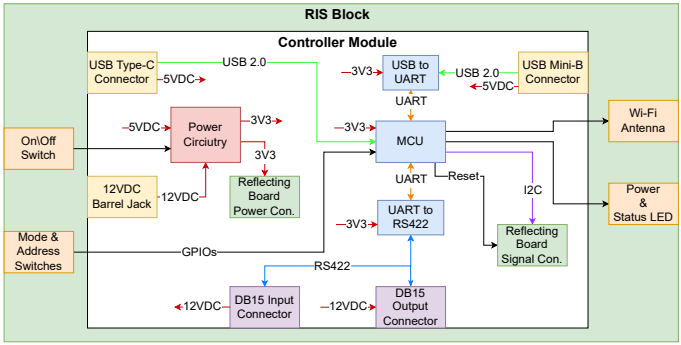}
    \caption{RIS controller module diagram}
    \label{fig:RIS Controller Module Diagram} \vspace{-10pt}
\end{figure}

The reflecting surface board is designed to receive configuration data and commands from the controller module and re-configures itself accordingly. The front surface of the board includes PIN diodes and features specially designed reflection patterns, while the rear surface contains integrated circuits controlling the PIN diodes, in addition to power and communication connectors for interfacing with the controller module. The controller module operates in two functional modes: master and slave. In master mode, the controller module is responsible for transmitting configuration data through a wired or wireless connection from a computer. It sends these data to the connected reflecting surface board and all slave modules in the system. In addition, it provides power distribution to all connected slave modules. In slave mode, the controller module receives the relevant configuration data from the master, configures the connected reflecting surface board accordingly, and forwards power to subsequent slave modules in the system. The overall operating diagram of the RIS controller module is shown in Fig. \ref{fig:RIS Controller Module Diagram}.

To facilitate self-configuration and inter-module communication, each module integrates an MCU supporting Wi-Fi wireless communication as well as I2C, UART, and USB interfaces. The MCU can wirelessly receive configuration data through Wi-Fi or a wired USB Type-C connection. A dedicated configuration application has been developed for the master module, which operates in conjunction with embedded software in the module. The I2C interface allows the module to load configuration data onto the connected reflecting surface board. The UART interface is converted to the RS422 protocol through integrated circuits, allowing bidirectional communication between blocks. Additionally, the USB Mini-B connector on the module is used for firmware updates and debugging of the MCU.

The module is equipped with two DB15 connectors, one input and one output, which facilitate the transmission of 12V DC power and RS422 communication signals between blocks. The controller module includes a toggle switch to select its operating mode (master or slave) and four additional toggle switches to address slave blocks. This configuration allows each block to be easily assigned an address and set to the appropriate mode during the RIS system installation.

\section{Experimental Evaluations}

\subsection{Evaluation of Unit Reflecting Element Response}
\begin{figure}[t]
\centering
    \includegraphics[width=0.5\textwidth]{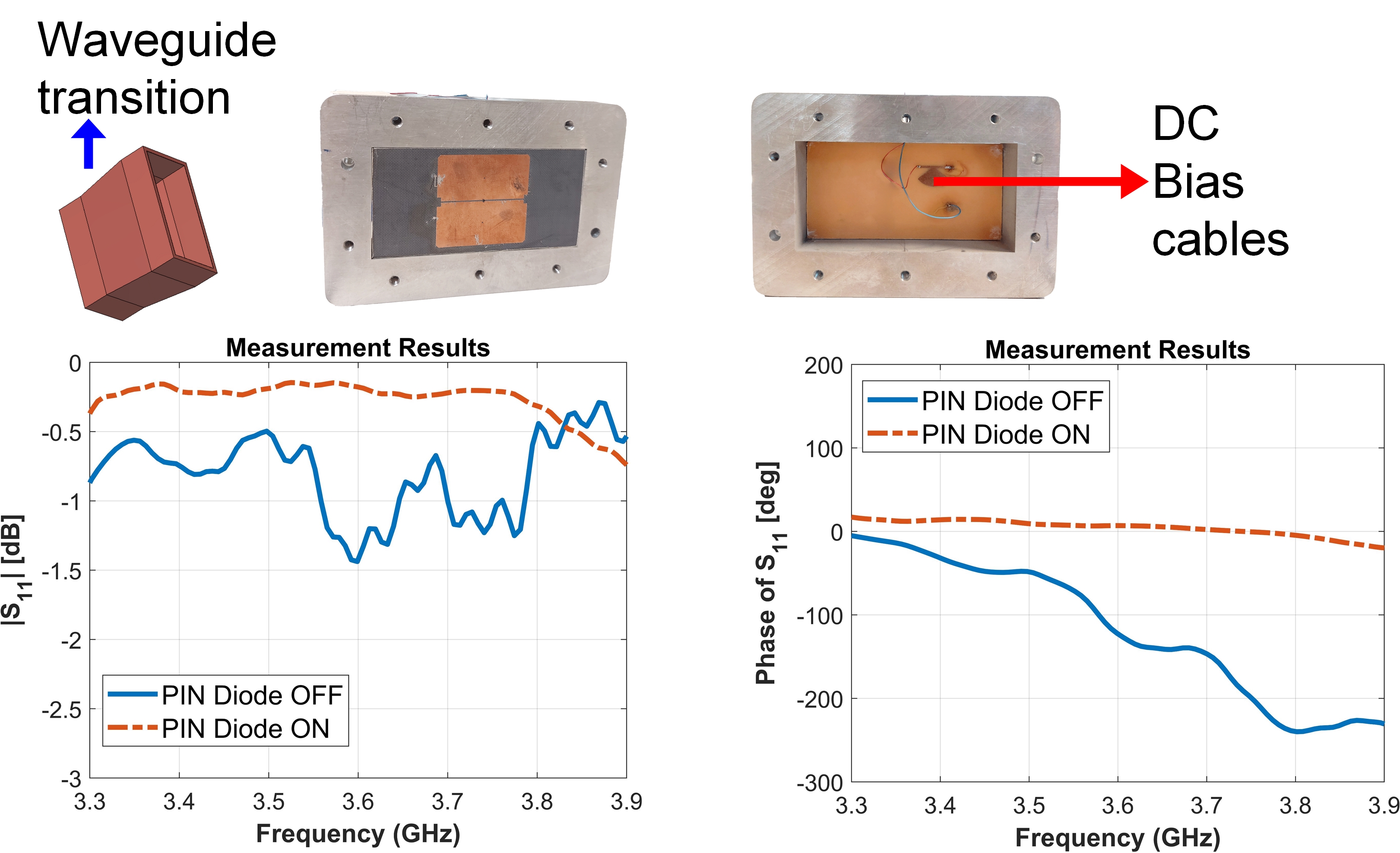}
\caption{Measurement results of reflection parameter (\(S_1\)\(_1\)) of the unit reflecting element inside waveguide transition } \vspace{-10pt}
\label{fig:ris-S11measurement}
\end{figure}

In order to measure reflection parameters (\(S_1\)\(_1\)), a small prototype of the unit reflecting cell is fabricated. A WR284 waveguide is chosen to measure \(S_1\)\(_1\) parameter of the structure. Since the sizes of the structure ($41\times 41$ mm) do not fit inside the WR284 waveguide, the authors have designed a waveguide extension to fit the structure and measured it, as in \cite{rana2021experimental}. The dimensions of the transition part length and extended waveguide apertures are $43$ mm and $82\times 41$ mm, respectively.
The waveguide transition structure and measurement setup are shown in Fig. \ref{fig:ris-S11measurement}. Firstly, the reflection is measured for $0$ V bias voltage for the OFF situation of the PIN diode. Then, the $0.85$ V bias voltage is applied to the unit reflecting element through the DC bias cables to switch the PIN diode from the OFF to the ON state. The measured reflection \(S_1\)\(_1\) parameter of the fabricated unit reflecting element is presented in  Fig. \ref{fig:ris-S11measurement}. The magnitude of the measured \(S_1\)\(_1\) parameters are over $-3$ dB for both cases, and the difference of unwrapped phase values between the ON and OFF states are almost in the range of $180 \pm 20$ degrees for $3.7$-$3.8$ GHz.

\subsection{RIS Block Performance}
\begin{figure}[t]
\centering
    \includegraphics[width = 0.35\textwidth]{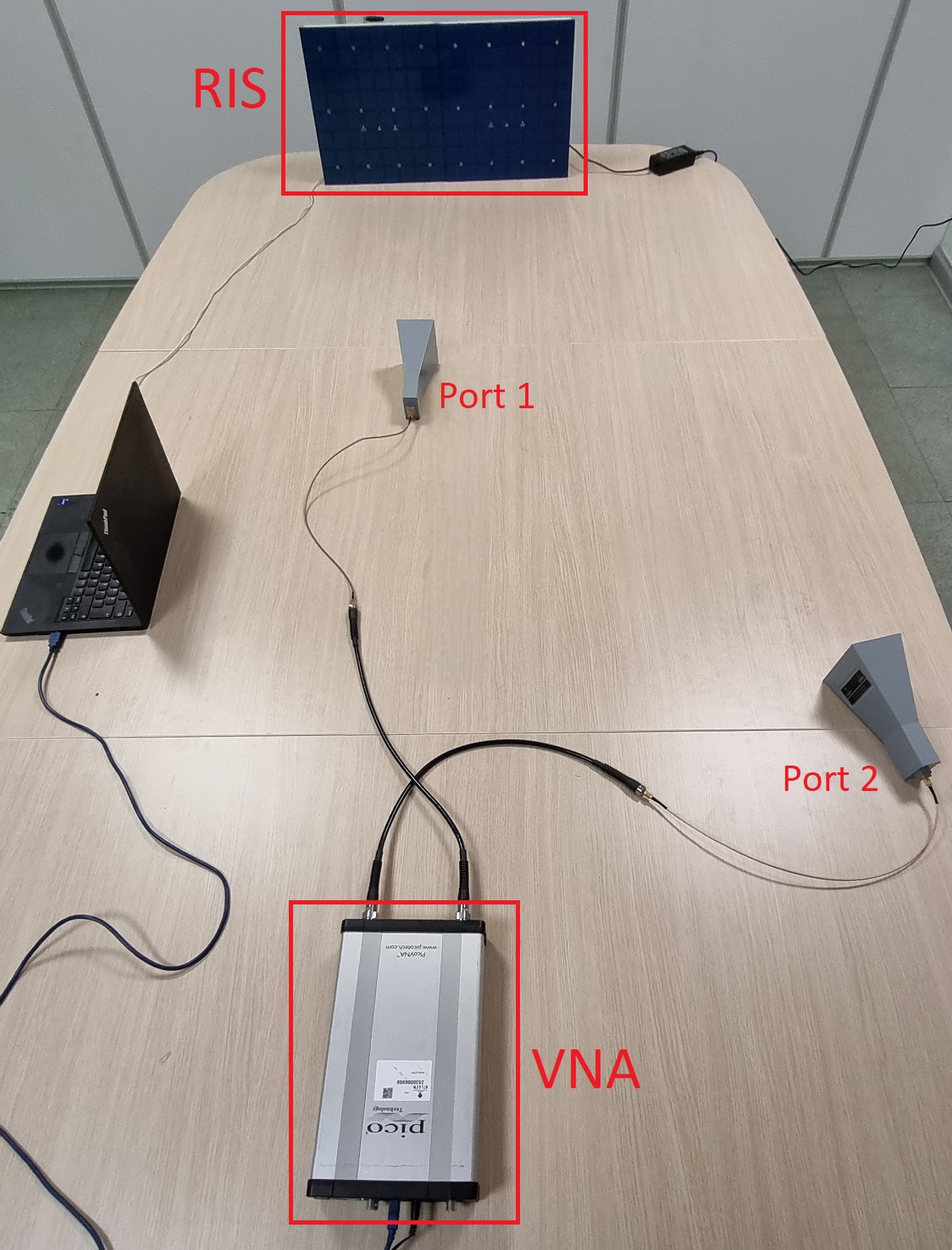}
\caption{The measurement setup} 
\label{fig:setup} \vspace{-10pt}
\end{figure}

\begin{figure}[t]
\centering
    \includegraphics[width = 0.23\textwidth]{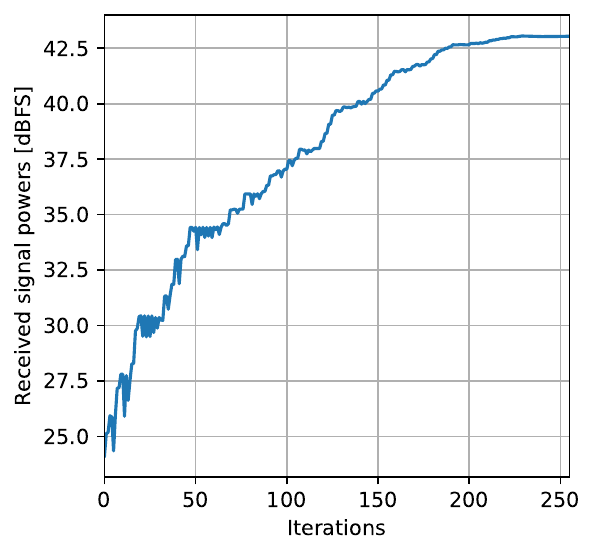}
    \includegraphics[width = 0.22\textwidth]{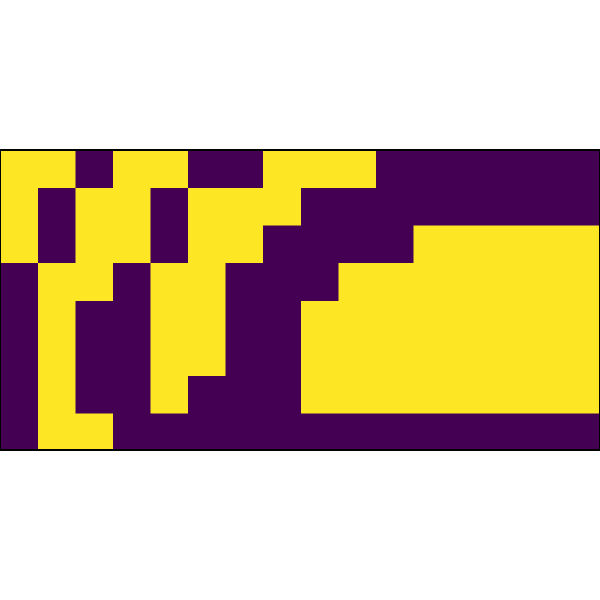} 
\caption{The received signal power using the iterative algorithm (left) and the final phase shift configuration (right)} \vspace{-10pt}
\label{fig:powers}
\end{figure}

Two RIS blocks are utilized, one in master and the other one in slave modes, in order to demonstrate the performance improvements. The measurement setup consisting of fabricated RIS, horn antennas, and vector network analyzer (VNA) is illustrated in Fig. \ref{fig:setup}. The transmitter and receiver horn antennas are positioned $1$ meter and $2$ meters away from the RIS, respectively, and at $0^\circ$ and $20^\circ$ angles relative to the RIS's normal, respectively. First, the iterative method \cite{kayraklik2024indoor} is utilized to determine the $8 \times 16$ reflecting element phase shift configuration, which focuses the incidence signal toward the receiver. Fig. \ref{fig:powers} shows the received signal power through the iterations, along with the resulting phase shift configuration of the RIS's reflecting elements. It can be observed from Fig. \ref{fig:powers} that the received signal power starts from around $24$ decibels relative to full scale (dBFS) and goes up to around $43$ dBFS. The improvement obtained from the RIS phase shift optimization is around $19$ dB for this placement of nodes. 

\begin{figure}[t]
\centering
    \includegraphics[width = 0.4\textwidth]{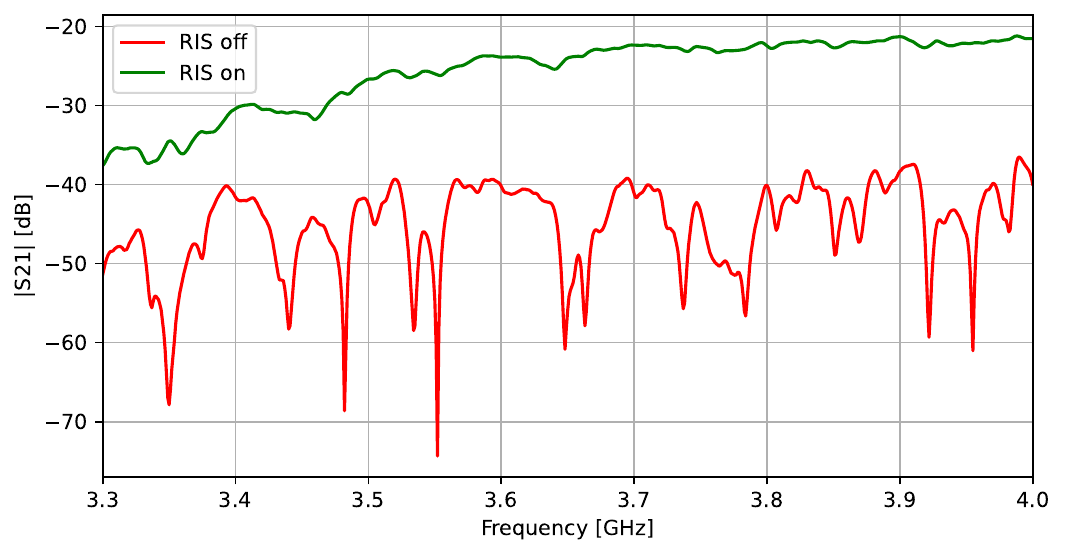}
\caption{The magnitude of \(S_2\)\(_1\) parameters} \vspace{-10pt}
\label{fig:S21VNA} 
\end{figure}

The performance of the RIS over different frequencies containing the n78 frequency band is validated for the given placement by measuring the forward voltage gain (\(S_2\)\(_1\)) using a VNA for observing reflections from the RIS. Fig. \ref{fig:S21VNA} shows the magnitude of \(S_2\)\(_1\) parameters for the frequencies starting from $3.3$ GHz to $4$ GHz. It can be observed from Fig. \ref{fig:S21VNA} that by applying the calculated RIS phase shift configuration, the \(S_2\)\(_1\) magnitudes are improved more than $15$ dB, and the RIS's performance degrades for lower frequencies. Note that the level of performance improvement depends on the placements of the nodes and may change from one setup to another.

\section{Conclusion}
In this paper, we designed and developed a PIN diode based RIS prototype operating at the n78 frequency band. The prototype has a modular architecture including one master and up to sixteen slave blocks, where an identical hardware structure with $8\times 8$ reflecting surface elements and a controller board can be configured as a master or a slave block. The states of the PIN diodes are set to ON or OFF by providing the appropriate voltage level through the controller board, providing a $180^\circ$ phase difference between the two states. The measurements demonstrated that the received signal power can be improved by more than $15$ dB for the given placement across the n78 frequency band using the iterative algorithm.


\section*{Acknowledgment}
The authors would also like to thank Aysun SAYINTI, Sertan SUVARI, and Mustafa KILIC for the waveguide measurements, as well as Sergiy B. PANIN for the phase matrices calculations at MILTAL and Bayram DENIZ for the PCB design at BILGEM.


\printbibliography

\end{document}